# A computational method for estimating Burr XII parameters with complete and multiple censored data

*S. Saei[a], M. Mohammadi[b], M. Fekri sari[c], K. Jenab[d]*
[a]Industrial Engineering Department, Science & Research Branch, Islamic Azad University, Tehran, Iran
[b]Department of Industrial Engineering, Mazandaran University of Science and Technology, Babol, Iran
[c]Industrial engineering Department, Qazvin Islamic Azad university, Tehran, Iran
[d]Society of Reliability Engineering, Ottawa, Canada

**Abstract**

Flexibility in shape and scale of Burr XII distribution can make close approximation of numerous well-known probability density functions. Due to these capabilities, the usages of Burr XII distribution are applied in risk analysis, lifetime data analysis and process capability estimation. In this paper the Cross-Entropy (CE) method is further developed in terms of Maximum Likelihood Estimation (MLE) to estimate the parameters of Burr XII distribution for the complete data or in the presence of multiple censoring. A simulation study is conducted to evaluate the performance of the MLE by means of CE method for different parameter settings and sample sizes. The results are compared to other existing methods in both uncensored and censored situations.

I. Introduction

The Burr XII distribution was introduced by Burr (1942) [1], and is becoming increasingly utilized in the contexts of lifetime data analysis, reliability analysis and actuarial science in order to reduce the likelihood of failure.

[1]Corresponding Author Email: saviz.saei@gmail.com

The probability density and cumulative distribution functions of Burr XII distribution are given by

$$f_X(x) = kcx^{c-1}(1+x^c)^{-(k+1)}, x \geq 0, c, k > 0 \qquad (1)$$

$$F_X(x) = 1 - (1+x^c)^{-k}, x \geq 0, c, k > 0 \qquad (2)$$

The useful applications of Burr XII distribution emanate from different curve shapes that it can expose. As is shown by [2] and [3], Burr XII distribution comprises the curve shape of normal, gamma, log-normal, logistic, exponential distributions as well as a part of Pearson Type I, II, V, IX and XII families. For instance, the normal density function may be approximated as a Burr XII distribution with $c$=4.85437 and $k$=6.22665 and the gamma distribution with shape parameter 16 can be approximated as a Burr XII distribution with $c$=3 and $k$=6, and the log-logistic distribution is a special case of the Burr XII distribution. The properties of Burr XII distribution are applied in the areas of quality control, economics, duration of failure time modeling. Zimmer and Burr (1963) developed a method for sampling variables from non-normal populations using the Burr XII distribution [4]. Dogru and Arslan[5] introduced the OBR estimation method for the Burr XII distribution. Marshall-Olkin Extended Burr XII (MOEBXII) distribution family, which is a generalization of Burr XII distribution proposed by Al-Saiari et al.[6] Brzeziski (2014) also suggested that the BXII distribution is useful for empirical distribution of journal impact factors[7].

Burr (1967) used his distribution to assess the effect of non-normality on the context of control chart for monitoring mean of the process [8]. Jones et al. (2014) applied it to modelling inpatient cost in English hospitals [9].

Chou et al. (2000) applied the Burr XII distribution to generate an economic statistical design of the control chart for the non-normally distributed data [10]. Doğru and Arslan [6,11] proposed an estimator based on M estimation and optimal B-robust (OBR) estimator to estimate the parameters of Burr XII distribution.

Abbasi (2009) used the Burr XII distribution in providing training data for a neural network to estimate the process capability index of non-normal processes [12]. Dogru et al. [13] describes robust estimators by using the OBR estimation method for the parameters of the generalized half-normal distribution. A study was conducted by Aslam et al. (2017) on a three-component mixture of Exponential, Rayleigh, Pareto, and Burr Type-XII distributions in relation to reliability analysis. This study aimed to derive algebraic expressions for different functions of survival time [14]. As noted by Tahir et al. (2015) the mixture of probability density functions from the same (different) family (families) known as Type-I (Type-II) mixture model. In many applications, the available data can be considered from a mixture of two or more distributions. This idea enables us to mix statistical distributions in order to get a new distribution [15]. Saadatmelli et al. (2018) developed a cost model based on the optimization of the average cost per unit of time. Indeed, the cost model compared assignable cause under the influence of a single match case with multiple assignable causes under a same cost and time parameters [16]. Noori-Asl et al. (2016) proposed the use of expectation-maximization (EM) algorithm to compute the maximum likelihood estimates(MLEs) of model parameters. Further, the asymptotic variance-covariance matrix of the MLEs derived by missed information principle can be utilized to construct asymptotic confidence intervals (CIs) for the parameters. The Bayes estimates of the unknown parameters were obtained under the assumption of gamma priors by using Lindley's approximation and Markov chain Monte Carlo (MCMC) technique [17]. Silva et al. (2008) presented an alternative model for unimodal shape of failure rate function as Log-Burr XII regression model with the presence of censored data. Due to this, the performance of this model compared to Log-Logistic Regression model based on variance, mean squared error and the

likelihood- ratio test [18]. Also Paranaiba et al. (2011) defined a five-parameter life time distribution, called beta Burr XII (BBXII) distribution. This model was useful for modeling lifetime data with unimodal-shaped hazard rate function [19]. Gunasekera (2018) proposed the reliability function of Burr XII distribution by the concept of generalized variable method based on progressively type II censoring with random removals, where the number of units removed at each failure time has a discrete uniform distribution [20]. Moore (2000) derived the reliability function for the two-parameter Burr type XII failure model under three different loss functions, absolute deference, squared error and logarithmic. In this paper, it is assumed that the parameter p behaves as a random variable having a gamma prior and a vague prior. Monte Carlo simulations are presented to show that the "popular" squared error loss function is not always the best. For larger sample sizes, the logarithmic loss function gives better estimates for p, however for smaller n, the squared error loss function presents better estimates to p [21].

Desirability of the Burr XII distribution can be waned by the difficulty of estimating its two parameters which are c and k. literally the successful applications of Burr XII distribution obtain when one could precisely estimate its parameters. However, considering method of moments to reach the Burr XII parameters forces to solve the following two nonlinear equations which would have more than one solution and it is not straightforward to obtain any (note that in using method of moments first we have to justify equation (4) to obtain the second order moment).

$$\mu_x = \frac{k}{\Gamma(k+1)} \Gamma\left(k - \frac{1}{c}\right) \Gamma\left(k + \frac{1}{c}\right), ck > 1 \tag{3}$$

$$\sigma_x = \left(\frac{k}{\Gamma(k+1)}\left(\Gamma\left(k - \frac{2}{c}\right)\Gamma\left(1 + \frac{2}{c}\right) - \frac{k\left(\Gamma\left(k - \frac{1}{c}\right)\Gamma\left(1 + \frac{1}{c}\right)\right)^2}{\Gamma(k+1)}\right)\right)^{1/2}, ck > 2 \tag{4}$$

where $\mu_x$ and $\sigma_x$ are the mean and standard deviation of the Burr XII distribution and can be replaced by mean and standard deviation of the sample. In some respects, when estimating parameters of a known family of the probability distributions, method of moments is superseded by the Maximum Likelihood Estimation (MLE) method, because the maximum likelihood estimators have higher probability of being close to the quantities to be estimated. Estimates by the method of moments may be used as the first approximation to the solutions of the likelihood equations. However, in some cases, infrequently with large samples but not so infrequently with small samples, the estimates given by the method of moments are outside of the parameter space; as these estimations are not necessarily sufficient statistics. That problem never arises when using MLE. Also, for a large class of problems, the MLE possesses a number of attractive asymptotic properties. Taking into account merits of MLE method, Wingo (1983) indicated that the MLE estimators for Burr XII distribution exist under a certain condition that at least one of the observations be less than one [22]. Practically this condition is easily satisfied as Burr XII has, for a wide range at its parameters values, a relatively big density mass below one. Also Wingo (1993) constructed likelihood function for censored data as well as complete data [23]. To use MLE method in estimating Burr XII parameters we step in maximizing a highly nonlinear function (log-likelihood function) given in equation (5) while by increasing the sample size, it turns out to be more difficult to maximize. Also equation (6) shows the log-likelihood function for multiple censoring situations with *r* out of *n* observations are censored.

$$Ln(L) = Ln\left(\prod_{i=1}^{n} f_X(x_i)\right) = n(Ln(c) + Ln(k)) + (c-1)\sum_{i=1}^{n} Ln(x_i) - (k-1)\sum_{i=1}^{n} Ln(1 + x_i^c) \qquad (5)$$

$$Ln(L_{r,n-r}) = Ln\left(\prod_{i=1}^{r} f_X(x_i) \prod_{j=r+1}^{n} (1 - F_X(x_i))\right) = \qquad (6)$$

$$\sum_{i=1}^{r}[Ln(k) + Ln(c) - (k+1)Ln(1+x_i^c) + (c-1)Ln(x_i)] - k\sum_{j=r+1}^{n} Ln(1+x_i^c)$$

Using the log-likelihood function, equation (6), the first order conditions for c and k for the case of complete data are given in equation (7) and (8), while equation (9) and (10) pertain to the case of censored data:

$$\frac{n}{c} - (k+1)\sum_{i=1}^{n}\left(\frac{x_i^c}{1+x_i^c}Ln(x_i)\right) + \sum_{i=1}^{n} Ln(x_i) = 0 \tag{7}$$

$$k = \frac{r}{\sum_{i=1}^{n} Ln(1+x_i^c)} \tag{8}$$

$$\frac{r}{c} - \sum_{i=1}^{r}\left(\frac{x_i^c}{1+x_i^c}Ln(x_i)\right) + \sum_{i=1}^{r} Ln(x_i) - k\sum_{j=r+1}^{r}\left(\frac{x_i^c}{1+x_i^c}Ln(x_i)\right) = 0 \tag{9}$$

$$k = \frac{r}{\sum_{i=1}^{n} Ln(1+x_i^c)} = 0 \tag{10}$$

An alternative is to solve equations (7) and (8) (or (9) and (10)) by the numerical methods such as Newton-Raphson method, however they are very sensitive to the initial values and also the solutions are not unique. Burr in 1942 constructed some standard tables to allow estimation of c and k based on the sample skewness and kurtosis. In these tables, for those skewness and kurtosis which are not in the tables, the curvilinear interpolation is used to find the parameters with desired moments. Recently, Abbasi et al. (2010) developed a neural network to estimate the Burr XII parameters with four parameters; this method mimics the concepts of method of moments as it estimates c and k by using the skewness and kurtosis of data [24]. An important advantage of the ANN method and using the Burr tables compared to method of moments and MLE is that it allows treating Burr XII distribution with four parameters. Wang and Cheng (2010) applied expectation-maximization (EM) method to estimate the Burr XII parameters for multiple censored data [25]. They compared EM method with Newton-Raphson (NR) algorithm in solving equations (9) and (10), and remarked that EM outperforms NR algorithm. In this study we develop Cross Entropy (CE) algorithm to maximize the log-likelihood of the Burr XII distribution for both complete and censored data. Then, we compare the CE method with the EM method for multiple censored data and the CE method with the ANN method for complete (uncensored) data.

The rest of the paper is organized as follows: Section 2 describes current method in estimating Burr XII distribution for both censored and uncensored data. In Section 3 a short explanation of Cross Entropy method is presented and Section 4 outlines steps of using CE in maximizing the log-likelihood function of Burr XII distribution. Test problems considering large number of replications appear in Section 5. Section 6 concludes the paper.

## II. Preliminaries

In this section we explain the ANN method for complete data and the EM algorithm for censored data. We will compare the results of the CE with these methods in Section 5.

### A. ANN method in estimating c and k

In the method of moments, the parameters are estimated based on the relationship between the moment of sample and the parameters. However, sometimes these relationships do not appear in simple equations. Relationship between moments and parameters in the Burr XII distribution is extraordinary complicated by using the concept of method of moments trained a neural network which is able to estimate the Burr XII parameters based on skewness and kurtosis calculated from the data [24]. They trained the neural network by using the Burr standard tables and came up with following closed formulas:

$$[c, \mu_z, \sigma_z] = Tansig(\omega_3 Tansig(\omega_2 Tansig(\omega_1 [Skewness(X), Kurtosis(X) - 3] + \beta_1) + \beta_2) + \beta_3) \quad (11)$$

$$k = \frac{n}{\sum_{i=1}^{n} Ln\left(1 + \left(\frac{(x_i - \mu_x)\sigma_z}{\sigma_x} + \mu_z\right)^c\right)} \quad (12)$$

in which $Tansig(n) = \frac{2}{(1+e^{-2n})} - 1$, $\mu_z$ and $\sigma_z$ are mean and standard deviation of the standard Burr XII variable and $\mu_x$ and $\sigma_x$ are mean and standard deviation of X, relationship between X and Z is expressed by $(X - \mu_x)/\sigma_x = (Z - \mu_z)/\sigma_z$. The parameters $\omega$'s and $\beta$'s are weights and bias values obtained in training process ($\omega$'s and $\beta$'s are presented in [19]. Abbasi et al. (2010) also provided some comments to improve the estimated parameters from neural network based method and used Monte Carlo simulations to show the performance behavior of ANN method [24]. We note that neural network based model is applicable when we have uncensored data.

*B. Expectation Maximization Algorithm*

The EM algorithm is an efficient iterative procedure to compute maximum likelihood estimation problems involving censored data, and it has been widely applied in parameter estimation of mixture model (see [26] and [27] for more details). Each iteration of the EM algorithm consists of two steps: The E-step, and the M-step. In the E-step, the missing (censored) data are estimated by using the conditional expectation given the observed data and current estimate of the model parameters. In the M-step, the likelihood function called Q-function is maximized under the assumption that actual value of censored data is known and current estimate of other parameters are used when maximizing Q-function respect to one of the parameters. Convergence is assured since the algorithm increase the likelihood at each iteration. We refer to a useful tutorial provided by [28] for more details on EM algorithm.

EM algorithm outlined for the Burr XII distribution with r complete and n- r censored data is:

E-step:

The probability density function for censored data when censoring occurs in time $d_j$ for $j = r, \ldots, n$ is given by:

$$f_X(x|x > d_j) = kc(1 + d_j^c)^k \frac{x^{c-1}}{(1 + x^c)^{k+1}} \quad x > d_j \tag{13}$$

then the Q-function is obtained by:

$$Q(c, k|c^{(m)}, k^{(m)}) = E\left(Ln(L(x, c, k))\right) = n(Ln(c) + Ln(k)) + (c - 1) \sum_{i=1}^{r} Ln(d_i)$$

$$-(k + 1) \sum_{i=1}^{r} Ln(1 + d_i^c) + (c - 1) \sum_{j=r+1}^{n} E(Ln(x_j)|x_j > d_j) \tag{14}$$

$$-(k + 1) \sum_{j=r+1}^{n} E(Ln(1 + x_j^c)|x_j > d_j)$$

M-step: in this step Q-function is maximized to estimated parameters in an iterative process, i.e.

$$(c^{(m+1)}, k^{(m+1)}) = argmax\left(Q(c, k|c^{(t)}, k^{(t)})\right) \tag{15}$$

Solving equation (14) is not straightforward as the last term of equation (14) includes unknown parameters c (see [25]). Wang and Cheng (2010) applied Taylor series expansion for the last term in the equation (13) and used Monte Carlo simulation to solve it [25].

### III. Cross-Entropy Algorithm

The cross-entropy algorithm originally was motivated by [29] for estimating probabilities of the rare events. Then it was modified to solve optimization problems [30]. Also, Moeini et al (2013) developed a cross-entropy based estimation method for the three parameter Weibull probability distribution [31]. They performed a comparative analysis among the current methods and showed their method is more efficient. CE method such as a generic and practical tool are applied for solving NP-hard problems [32, 33, 34, 35].

The CE based algorithm is an iterative procedure which involves in two phases:

1. Generate a random data sample according to a specified mechanism.

2. Update the parameters of the random mechanism based on the data to produce a better sample in the next iteration.

There is a wealth of literature on theory and an application of CE, however presenting the detail literature is not in the scope of this paper. The CE method homepage can be found at: http://www.cemethod.org/.

Kroese et al. (2006) presented the effectiveness of the CE method for solving difficult continuous optimization problems. In this regard, a family of pdfs $\{f(.,v), v \in V\}$ on set X defines to maximize function $S(x), x \in X$. Let $\gamma^* = \max_{x \in X} S(x)$, by taking a random sample $X_1,...,X_N$ that it can estimate

$$\ell(\gamma) = Pr(S(x) \geq \gamma | X \sim f(.,v), v \in V) = E(I_{\{S(x) \geq \gamma\}} | X \sim f(.,v), v \in V) \tag{16}$$

Noting that $S(x) \geq \gamma$ is a rare event when $\gamma$ is set to $(1-\varrho)^{th}$ quantile of $S(x)$ where $(1-\varrho)^{th}$ is close to 1. By taking an efficient strategy to update $v$, the CE method generates a sequence of tuples $\{(\gamma_t, v_t)\}$, which converges promptly to a small neighborhood of the optimal tuple $(\gamma^*, v^*)$ [36].

$v$ is updated in an iterative procedure. In each iteration $v$ is updated based on the current taken sample from $f(.,v_{t-1})$ i.e. $x_1^{(t-1)},...,x_N^{(t-1)}$N (sample in iteration $(t-1)^{th}$)) by using following equations. Equation (17) is applied and then parameters are smoothed by equation (18).

$$\tilde{v}_t = \underset{v}{argmax}\left(\frac{1}{N}\sum_{i=1}^{N} I_{\{S(x_i^{(t-1)}) \geq \hat{\gamma}_{t-1}\}} \ln f(x_i^{(t-1)}, v)\right) \tag{17}$$

$$\hat{v}_t = \alpha \tilde{v}_t + (1-\alpha)\tilde{v}_{t-1} \tag{18}$$

Where $\hat{\gamma}_i$ i is $(1-\varrho)^{th}$ quantile of $S(x)$ obtained from $\left(x_1^{(t-1)}, \ldots, x_N^{(t-1)}\right)$ and $\alpha$ is called smoothing parameter and chosen from [0.7, 1). Iteration will proceed to define termination condition.

In typical applications, (17) is a convex function and differentiable with respect to $v$, see also [37] Thus, the solution of (17) may be readily obtained by solving the following system of equations:

$$\frac{1}{N}\sum_{i=1}^{N} I_{\{S(x_t^{(t-1)}) \geq \hat{\gamma}_{t-1}\}} \nabla \ln f(x_i^{(t-1)}, v) = 0 \tag{19}$$

where the gradient is with respect to $v$. The advantage of this approach is the solution of (19) that can often be calculated analytically. In particular, this happens if the distributions of the random variables belong to a natural exponential family. For further details, we refer to [38].

## IV. MLE via CE for Burr XII

The steps of CE in optimizing MLE function of Burr XII distribution for both complete and multiple censored data are same, but as noted earlier the likelihood functions are different. In implementing CE for estimate the Burr XII parameters ($c$ and $k$), we choose the left-truncated normal distribution for generating solutions, therefore:

$$c_t \sim N_{left-trancated}(\widehat{\mu_c}_{t-1}, \widehat{\sigma_c}_{t-1}^2) \text{ and } k_i \sim N_{left-trancated}(\widehat{\mu_k}_{t-1}, \widehat{\sigma_k}_{t-1}^2)$$

Initial parameters for mean and standard deviation of the left-truncated normal distributions are set to 0 and 10 respectively for both $c$ and $k$. Let $v = [\mu_c, \mu_k, \sigma_c, \sigma_k]$, solving equation (17) for this application considering the truncated normal distribution for $\{f(.,v), v \in V\}$ gives the following updating equations in each iteration:

$$\widehat{\mu_c}_t = \sum_{i \in \mathcal{L}} c_{i,t-1}/N^{elite}, \quad \widehat{\mu_k}_t = \sum_{i \in \mathcal{L}} k_{i,t-1}/N^{elite}$$

$$\hat{\sigma}_{c_t}^{\,2} = \sum_{i \in \mathcal{L}} (c_{i,t-1} - \widehat{\mu}_{c_t})^2 \Big/ N^{elite}, \qquad \hat{\sigma}_{k_t}^{\,2} = \sum_{i \in \mathcal{L}} (k_{i,t-1} - \widehat{\mu}_{k_t})^2 \Big/ N^{elite}$$

Where $L$ is the set of indices of the best solutions $(c_i, k_i)$ in each iteration, $N^{elite}$ is the number of best selected solutions. Let $N^{elite} = N\varrho$, and $\varrho$ is the quantile value which we set it to 0.1. We use equation (18) to smooth the parameters $\widehat{\mu}_{c_t}, \widehat{\mu}_{k_t}, \hat{\sigma}_{c_t}$ and $\hat{\sigma}_{k_t}$, where smoothing factor are chosen $\alpha = 0.8$ for $\widehat{\mu}_{c_t}$ and $\widehat{\mu}_{k_t}$ and $\beta = 0.6$ for $\hat{\sigma}_{c_t}$ and $\hat{\sigma}_{k_t}$. These two factors are employed to prevent converging towards local optimal solutions. The stopping condition in our implementation is depended on $\hat{\sigma}_{c_t}$ and $\hat{\sigma}_{k_t}$ so, we stop the algorithm when $\max(\hat{\sigma}_{c_t}, \hat{\sigma}_{k_t})$ is less than a threshold $\varepsilon = 0.005$.

The steps of CE method in estimating the Burr XII parameters also is outlined in Algorithm 1.

**Algorithm 1. General CE algorithm for Burr XII parameter estimation**

Step 1: **Initialize:** Choose $\widehat{\mu}_{c_0}, \hat{\sigma}_{c_0}$ and $\widehat{\mu}_{k_0}, \hat{\sigma}_{k_0}$, Set $t=0$.

Step 2: **Repeat:**
Step 3: **Draw:** Increase t by 1. Generate a random vector sample $x_1, \ldots, x_N$ from the:
$c_i \sim N(\widehat{\mu}_{c_{t-1}}, \hat{\sigma}_{c_{t-1}}^{\,2})$ and $k_i \sim N(\widehat{\mu}_{k_{t-1}}, \hat{\sigma}_{k_{t-1}}^{\,2})$ distribution.
Where $x_i = (c_i, k_i) \quad i = 1, \ldots, N$.

Step 4: **Select:** Let $\mathcal{L}$ be the indices of the $N^{elite}$ best performing (=elite) samples.
Step 5: **Update:** Via the following formula:

$$\widetilde{\mu}_{c_t} = \sum_{i \in \mathcal{L}} c_i \Big/ N^{elite}, \qquad \widetilde{\mu}_{k_t} = \sum_{i \in \mathcal{L}} k_i \Big/ N^{elite}$$

$$\widetilde{\sigma}_{c_t}^{\,2} = \sum_{i \in \mathcal{L}} (c_i - \widetilde{\mu}_{c_t})^2 \Big/ N^{elite}, \qquad \widetilde{\sigma}_{k_t}^{\,2} = \sum_{i \in \mathcal{L}} (k_i - \widetilde{\mu}_{k_t})^2 \Big/ N^{elite}.$$

Step 6: **Smooth:**
$$\widehat{\mu}_{c_t} = \alpha \cdot \widetilde{\mu}_{c_t} + (1-\alpha)\widehat{\mu}_{c_{t-1}}, \quad \widehat{\mu}_{k_t} = \alpha \cdot \widetilde{\mu}_{k_t} + (1-\alpha)\widehat{\mu}_{k_{t-1}}$$
$$\hat{\sigma}_{c_t} = \beta \cdot \widetilde{\sigma}_{c_t} + (1-\beta)\hat{\sigma}_{c_{t-1}}, \quad \hat{\sigma}_{k_t} = \beta \cdot \widetilde{\sigma}_{k_t} + (1-\beta)\hat{\sigma}_{k_{t-1}}.$$

Where $\alpha$ and $\beta$ $\alpha$ are fixed smoothing parameter chosen from the interval [0, 1].

Step 7: **Until:** max $(\widehat{\sigma}_{c_t}, \widehat{\sigma}_{k_t}) < \varepsilon$

To implement the CE method for some problems, one might need to generate samples in some special regions due to the constraints of the underlying problem. For such random vector generation methods, we refer the interested readers to [39, 40, 41].

## V. Simulation and Comparison Studies

In this section we utilize the CE algorithm in estimating the Burr XII parameters with different parameters. Moreover, we compare the result in terms of precision and accuracy with the ANN method for complete data and the EM algorithm for multiple censored data. All of the simulations are implemented in MATLAB R2009b on a PC equipped with INTEL 2 Core, 1.83-GHz CPU and 512 MB RAM memory.

### A. Complete Data

We use inverse-method to generate the Burr XII data by using $x = \left(u^{-1/k} - 1\right)^{1/c}$, ($u$ is a random number drawn from Uniform (0, 1)). In the simulation study different sample sizes $n = 50, 100, 1000, 2500$ and $10000$ and different values of $c$ and $k$ are considered. The result of simulation runs is presented in Table 1. Using 1000 replications the mean and the standard deviation of estimated parameters are displayed for both CE and ANN methods. Moreover, CPU times of both methods are reported. The results reveal that the CE method outperforms in all cases in terms of accuracy and precision. Accuracy is defined by $|\hat{\theta} - \theta|$ where $\theta$ is real value of the parameter and $\hat{\theta}$ is its estimated value and precision is the standard deviation of $\hat{\theta}$. It is noteworthy that a comprehensive comparison study between EM Algorithm and Bayesian method for estimation of parameters of Burr XII distribution on censoring data is considered in [42]. In this study, we have no prior information on the unknown parameters, then it is always preferable to use the EM algorithm rather than the Bayes estimators, because the Bayes estimators are computationally more expensive.

Table 1: Burr XII parameters estimation for the complete data in 1000 replications

| Methods | | CE via MLE | | | | | ANN Method* | | | |
|---|---|---|---|---|---|---|---|---|---|---|
| Parameters | n | $\hat{c}$ | | $\hat{k}$ | | CPU | $\hat{c}$ | | $\hat{k}$ | |
| | | Mean | Std | Mean | Std | Time(s) | Mean | Std | Mean | Std |
| C=2 And K=5 | 50 | 2.0533 | 0.2186 | 5.32 | 0.9953 | 2.8719 | 1.6065 | 0.3896 | 4.205 | 1.1816 |
| | 100 | 2.0229 | 0.1459 | 5.1256 | 0.6437 | 3.1089 | 1.6088 | 0.3532 | 4.2142 | 1.0591 |
| | 1000 | 2.0057 | 0.0455 | 5.0092 | 0.192 | 7.5496 | 1.8145 | 0.2462 | 4.5953 | 0.5782 |
| | 2500 | 2.0003 | 0.0277 | 5.0029 | 0.1233 | 15.7474 | 1.8841 | 0.2483 | 4.7489 | 0.5737 |
| | 10000 | 1.9993 | 0.0139 | 4.9978 | 0.0624 | 54.8257 | 1.9586 | 0.3637 | 4.9288 | 0.5545 |
| C=3 And K=4 | 50 | 3.0692 | 0.3376 | 4.1699 | 0.708 | 2.7276 | 2.2976 | 0.9606 | 3.5202 | 1.8816 |
| | 100 | 3.0473 | 0.2356 | 4.103 | 0.4872 | 2.9665 | 2.5345 | 0.8094 | 3.5291 | 1.8765 |
| | 1000 | 3.0071 | 0.0694 | 4.0074 | 0.1389 | 7.1485 | 2.8966 | 0.5113 | 3.8751 | 0.9341 |
| | 2500 | 3.0013 | 0.0433 | 4.0027 | 0.0897 | 14.8678 | 2.8988 | 0.3879 | 3.8987 | 0.5305 |
| | 10000 | 3.0001 | 0.0212 | 4.0019 | 0.0435 | 51.9288 | 2.9576 | 0.302 | 3.9662 | 0.4989 |
| C=4.14 And K=9.13 | 50 | 4.2355 | 0.431 | 9.8791 | 2.3758 | 3.9443 | 3.9209 | 2.6531 | 10.1662 | 12.7621 |
| | 100 | 4.1997 | 0.3109 | 9.5596 | 1.5998 | 3.7934 | 3.9113 | 2.6477 | 10.1661 | 12.5619 |
| | 1000 | 4.1484 | 0.0931 | 9.1861 | 0.4196 | 8.4444 | 4.0531 | 0.5963 | 9.0436 | 2.3434 |
| | 2500 | 4.1459 | 0.0582 | 9.165 | 0.2822 | 16.9707 | 4.0758 | 0.3831 | 9.061 | 1.3703 |
| | 10000 | 4.1427 | 0.03 | 9.1329 | 0.1396 | 57.9248 | 4.0998 | 0.1846 | 9.0781 | 0.6427 |

*The CPU time for ANN method is less than 0.05 second for all cases

The results of Table 1 are also depicted in Figures 1. For both ANN and CE methods, the standard deviations of $\hat{c}$ and $\hat{k}$ always decrease when $n$ increases. Likewise, the $|\hat{\theta} - \theta|$ is decreasing in $n$, expect for one case. The $|\hat{\theta} - \theta|$ for $c$ in the CE method where $n$ goes from 2500 to 10000, goes from 0.0003 to 0.0007, the reason is that when sample size increase to 10000, optimizing the likelihood functions becomes more difficult and it increases the chance that the CE method traps in a local optimum. The CPU time for ANN method is very small and less than consumed time for the CE method in all cases as the ANN method just requires estimating skewness and kurtosis of the data and computes equation (11). However, $\hat{c}$ and $\hat{k}$ obtained by the CE method have more accuracy (less bias) and higher precision (less standard deviation) than ANN method.

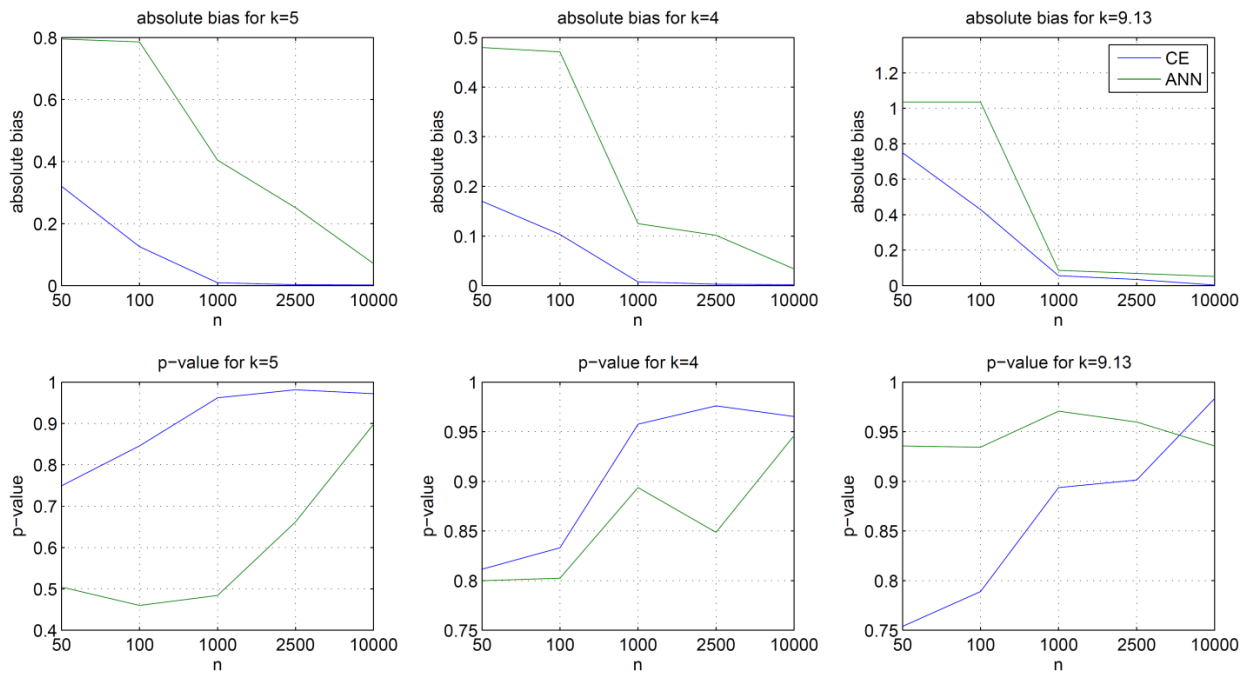

Figure 1: The absolute bias of estimation parameters for the complete data

*B. Multiple Censored Data*

For multiple censored data we apply the CE method and the EM algorithm for different parameter settings of Burr XII distributions considering different sample sizes. In presence of multiple censored data, CL is the level of censoring and 100 CL% indicates percentage of the data which are censored, to generate random data required in simulation study we employ the algorithm used by Wang and Cheng (2010) which is:

1- Generate *n* random data $(y_1, \ldots, y_n)$ from the Burr XII distribution with defined parameters

2- Set $r = n(1 - CL)$ to be the number of complete data. Then $n - r$ is the number of censored data

3- Let

$$\omega_i = \begin{cases} 1 & i = 1, ..., r \\ u_i & i = r+1, ..., n \end{cases}$$

Where $u_i$ is drawn from Uniform (0,1).

4- Set $x_i = y_i \omega_i$, $i = 1, ..., n$. Now, $(x_1, ..., x_n)$ becomes a multiple censored data set.

The simulation results are shown in Table 2 for CL=0.2 and in Table 3 for CL=0.6. Using 1000 replications for each case, the mean and the standard deviation of the estimated parameters for both the CE and EM methods are shown in Tables 2 and 3. Moreover, CPU time of the CE method is presented.

The results of EM algorithm are taken from Wang and Cheng (2010). Figures 2 and 3 depict the results of Tables 2 and 3 respectively. Figures 2 and 3 show that in both the CE and EM methods sometimes $|\hat{\theta} - \theta|$ increases by increasing the sample sizes, it is due to the impact of censoring, as when sample size increases the number of censored data also increases. However, the standard deviations of $c$ and $k$ always decrease by increasing the sample size in both methods.

The results indicate that in terms of accuracy $|\hat{\theta} - \theta|$ in the most of cases the CE method gives better results compare to the EM algorithm, this is more highlighted for larger value of $c$ and $k$. In terms of precision in some cases $\hat{c}$ and $\hat{k}$ obtained by the EM algorithm has less standard deviation. However, when there is a significant bias in estimation, small standard deviation is not a remarkable advantage of an estimator; this is true for the most of the cases which the EM algorithm gives a smaller standard deviation. The CPU time for the EM method is very small and less than the consumed time of CE method in all cases.

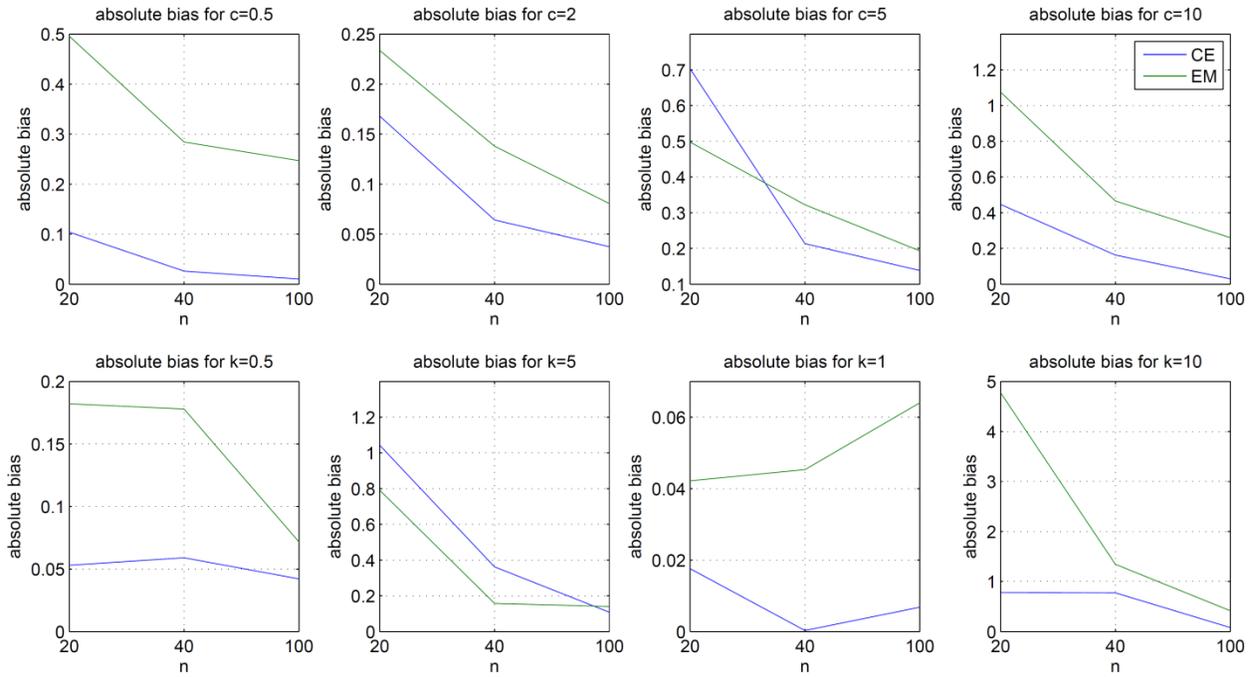

Figure 2: The absolute bias of estimation parameters for the censored data (CL = 0:2)

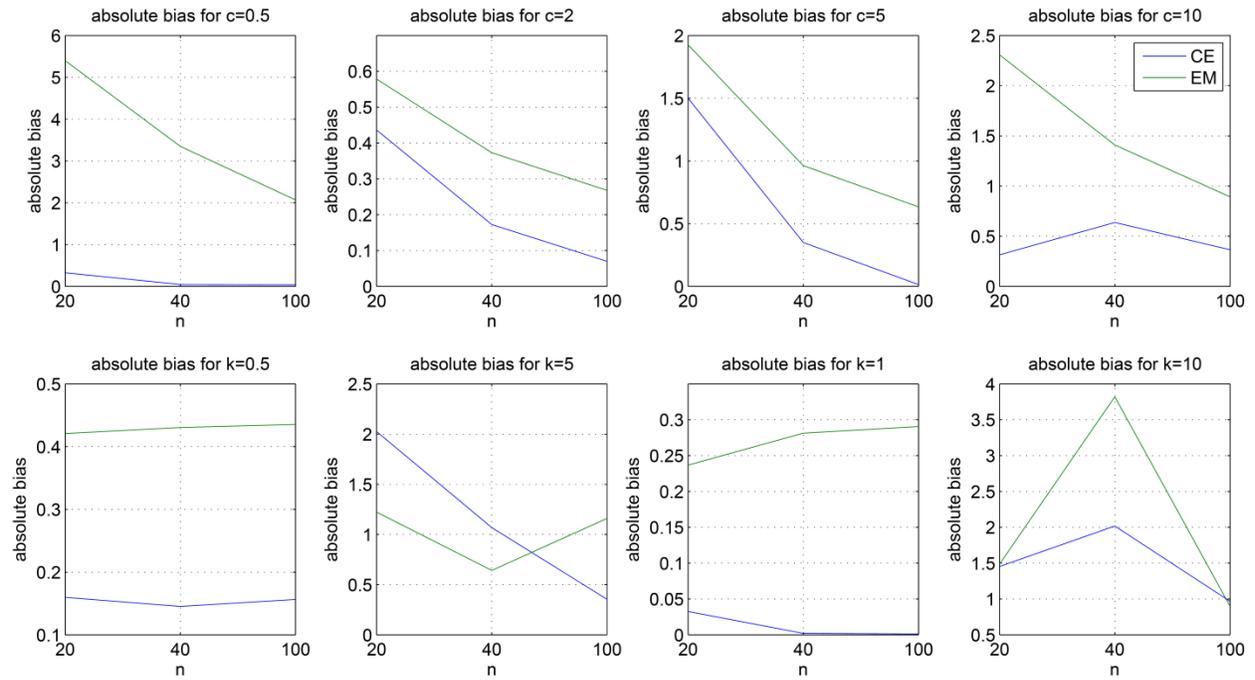

Figure 3: The absolute bias of estimation parameters for the censored data (CL = 0:6)

Table 2: Burr XII parameters estimation for the censored data (CL = 0:2) in 1000 replications

| Methods | | CE via MLE | | | | | EM Method* | | | |
|---|---|---|---|---|---|---|---|---|---|---|
| Parameters | N | $\hat{c}$ | | $\hat{k}$ | | CPU Time(s) | $\hat{c}$ | | $\hat{k}$ | |
| | | Mean | Std | Mean | Std | | Mean | Std | Mean | Std |
| C=0.5 | 20 | 0.604 | 0.2508 | 0.447 | 0.1629 | 2.4786 | 0.9957 | 1.2872 | 0.3179 | 0.159 |
| And | 40 | 0.5261 | 0.1275 | 0.441 | 0.1124 | 2.5105 | 0.7843 | 0.3102 | 0.3221 | 0.1182 |
| K=0.5 | 100 | 0.4897 | 0.0646 | 0.4579 | 0.0689 | 2.74 | 0.747 | 0.1797 | 0.4283 | 0.0882 |
| C=2 | 20 | 2.168 | 0.4161 | 6.0425 | 2.6038 | 2.5705 | 2.2337 | 0.4214 | 5.7907 | 2.7288 |
| And | 40 | 2.0642 | 0.293 | 5.3622 | 1.2711 | 2.6423 | 2.1379 | 0.2779 | 5.1577 | 1.3334 |
| K=5 | 100 | 2.0374 | 0.1683 | 5.1083 | 0.7024 | 2.8934 | 2.0806 | 0.1627 | 4.8596 | 0.6984 |
| C=5 | 20 | 5.7028 | 1.8154 | 1.0176 | 0.3274 | 2.5094 | 5.4976 | 1.6079 | 0.9578 | 0.3219 |
| And | 40 | 5.2135 | 1.2436 | 0.9997 | 0.2121 | 2.5628 | 5.3223 | 0.8735 | 0.9547 | 0.198 |
| K=1 | 100 | 5.1384 | 0.5666 | 1.0068 | 0.1248 | 2.842 | 5.193 | 0.5539 | 0.936 | 0.1144 |
| C=10 | 20 | 10.4466 | 1.6342 | 10.7806 | 2.9531 | 3.3105 | 11.0748 | 2.3548 | 14.7755 | 11.9488 |
| And | 40 | 10.1631 | 1.1019 | 10.7745 | 2.4558 | 3.2275 | 10.4658 | 1.3972 | 11.3443 | 3.8291 |
| K=10 | 100 | 10.0298 | 0.7311 | 10.0808 | 1.5612 | 3.1505 | 10.2602 | 0.8233 | 10.4165 | 1.938 |

*The CPU time for EM method is less than 1 second for all cases

Table 3: Burr XII parameters estimation for the censored data (CL = 0:6) in 1000 replications

| Methods | | CE via MLE | | | | | EM Method* | | | |
|---|---|---|---|---|---|---|---|---|---|---|
| Parameters | N | $\hat{c}$ | | $\hat{k}$ | | CPU Time(s) | $\hat{c}$ | | $\hat{k}$ | |
| | | Mean | Std | Mean | Std | | Mean | Std | Mean | Std |
| C=0.5 | 20 | 0.8261 | 1.6271 | 0.3403 | 0.1757 | 2.6709 | 5.8994 | 10.91 | 0.0792 | 0.0865 |
| And | 40 | 0.5444 | 0.4527 | 0.3548 | 0.1346 | 2.6508 | 3.85 | 5.8888 | 0.0695 | 0.0595 |
| K=0.5 | 100 | 0.4628 | 0.0941 | 0.3437 | 0.0792 | 2.8211 | 2.5668 | 1.3092 | 0.0647 | 0.0448 |
| C=2 | 20 | 2.4363 | 0.7035 | 7.0248 | 3.8824 | 2.7097 | 2.5782 | 0.763 | 6.2229 | 9.6315 |
| And | 40 | 2.1725 | 0.4551 | 6.0681 | 2.1453 | 2.6431 | 2.3729 | 0.4363 | 4.3582 | 1.9011 |
| K=5 | 100 | 2.0699 | 0.2395 | 5.3536 | 1.0363 | 2.8413 | 2.2677 | 0.2647 | 3.8396 | 0.8379 |
| C=5 | 20 | 6.4999 | 2.91 | 0.9675 | 0.3522 | 2.6186 | 6.9254 | 5.7298 | 0.7633 | 0.3563 |
| And | 40 | 5.3498 | 1.4501 | 0.998 | 0.2606 | 2.6016 | 5.9618 | 1.5888 | 0.7187 | 0.203 |
| K=1 | 100 | 5.0142 | 0.6654 | 0.9988 | 0.1849 | 2.8406 | 5.6335 | 0.8697 | 0.7095 | 0.1175 |
| C=10 | 20 | 10.3138 | 2.0805 | 11.4532 | 4.1205 | 3.7239 | 12.3066 | 3.5607 | 11.4873 | 23.1228 |
| And | 40 | 10.6371 | 1.4947 | 12.0174 | 3.2077 | 3.8583 | 11.4086 | 2.2252 | 13.8203 | 10.7177 |
| K=10 | 100 | 10.3648 | 1.0154 | 10.9664 | 2.5792 | 3.5775 | 10.8918 | 1.3881 | 10.8961 | 3.7214 |

*The CPU time for EM method is less than 1 second for all cases

## VI. An Illustrative Examples

Carbon Fiber Reinforced Plastic (CFRP) materials are vastly used in industries because of their mechanical properties. The fiber angle has significant impact on the strength and life span of the CFRP. As shown in Figure 4, Multiaxial Warp Knitted Fabrics can be used to create a type of multidirectional continuous fiber composite with angles -45° versus the filing 90°. They are composed of one to four layers of parallel fibers arranged at different orientations from each other, then stitched together with a thread that is made of polyester, forming a fabric. Here, the stitching pattern is known as "warp knitting" because many separate pieces of the thread, also known as yarn, runs in a zig-zag or straight pattern along the fabric, interlinked in the form of a chain on the back side of the fabric. If the fiber is arranged along the zig-zag pattern, this is called the warp (0°) direction, versus the filling direction (90°).

Figure 4: Schematic of multiaxial warp knitting processing

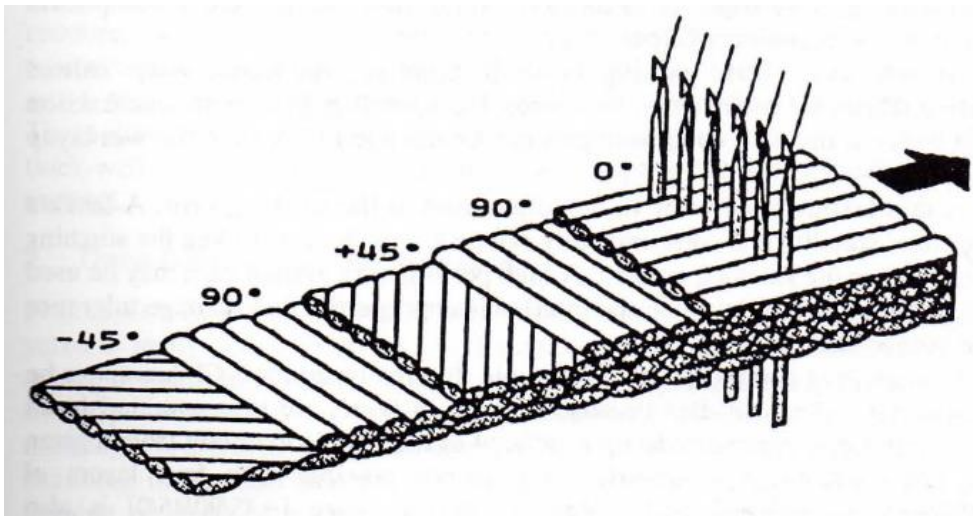

The following rounded data has been gathered form pieces of CFRP laminate with mixed fiber angles in Table 4. The distribution of these angles are found through analyzing images of laminate samples collected from optical microscopy.

Table 4: Compressing Testing Results

| Sample | Compressing force | (Angle, No of sample) |
|---|---|---|
| 5×Coupon 1 | 500 MPa | (-45,4), (-60,1) |
| 5×Coupon 2 | 600 MPa | (90,2), (45,4) |
| 5×Coupon 3 | 610 MPa | (45,1), (-30,3), (+60,1) |
| 5×Coupon 4 | 580 MPa | (-60,3), (+60,2) |
| 5×Coupon 5 | 590 MPa | (-45,1), (90,3), (-60,1) |
| 5×Coupon 6 | 580 MPa | (+606,3), (-30,2) |
| Angles: 1)-45, 2)90 3)-60 4)45 5)-30 6)+60 | | |

MATLAB's distribution fitting GUI and K-S (Kolmogorov-Smirnov) test has been used to calculate R. As Table 5 illustrates, the Burr PDF with $c=4.9$ and $k=6.3$ that are estimated with the proposed method has the highest p-value. Therefore, it should be taken as the best fit. The extreme value PDF is the second highest p-value with $\mu= 591.367$ and $\sigma= 21.5514$. If the function is log-logistic, then $\mu= \exp(a)$, $\sigma=\exp(b)$, where $a$ and $b$ are the values found in MATLAB. Fitting the data in Table 4 would result in the following p-values:

Table 5: P-values for Fitted PDF's for Compression Testing Results

| Distribution | p-value |
|---|---|
| Burr Value | 0.7812 |
| Extreme Value | 0.6458 |
| Generalized Extreme Value | 0.3537 |
| Inverse Gaussian | 0.2720 |
| Log-logistic | 0.6162 |
| Lognormal | 0.2847 |
| Normal | 0.3131 |

## VII. Conclusion

In this paper we developed a Cross-Entropy algorithm in estimating the Burr XII parameters for complete and multiple censored data. We conducted simulation studies to evaluate and compare the proposed method with the ANN method which is used in estimating the Burr XII parameters for complete data and Expectation Maximization algorithm which is suggested recently in estimating the Burr XII parameters for multiple censored data. The simulations results pointed out that the CE method outperforms the ANN method and performs better than the EM algorithm in most of cases in terms of the absolute value of bias in estimations. It should be noted that ANN method can be treated as estimating the Burr XII parameters with four parameters which still remains as an advantage of the ANN.

The results in this paper recommend the CE algorithm for estimating parameters of other distributions with complicated likelihood function.

## Acknowledgment

Authors would like to express their sincere appreciation to the editor and anonymous referees for their valuable comments that help us to improve the quality of the paper.